\documentclass[twocolumn,prb,showpacs,amsmath,amssymb]{revtex4}
\usepackage[dvips]{graphicx}
\usepackage{float}
\begin{document}

\title{Terahertz conductivity of the heavy-fermion compound UNi$_2$Al$_3$}

\author{Julia P. Ostertag}
\author{Marc Scheffler}
\email[*]{marc.scheffler@pi1.physik.uni-stuttgart.de}
\author{Martin Dressel}
\affiliation{1. Physikalisches Institut, Universit\"at Stuttgart, Pfaffenwaldring 57, 70550 Stuttgart, Germany}
\author{Martin Jourdan}
\affiliation{Institut f\"ur Physik, Johannes Gutenberg-Universit\"at Mainz, Staudingerweg 7, 55128 Mainz, Germany}

\date{\today}

\begin{abstract}
We have studied the optical properties of the heavy-fermion compound UNi$_2$Al$_3$ at frequencies between 100~GHz and 1~THz (3~cm$^{-1}$ and 35~cm$^{-1}$), temperatures between 2~K and 300~K, and magnetic fields up to 7~T. From the measured transmission and phaseshift of radiation passing through a thin film of UNi$_2$Al$_3$, we have directly determined the frequency dependence of the real and imaginary parts of the optical conductivity (or permittivity, respectively). At low temperatures the anisotropy of the optical conductivity along the {\textit a}- and {\textit c}-axes is about 1.5. The frequency dependence of the real part of the optical conductivity shows a maximum at low temperatures, around 3~cm$^{-1}$ for the $a$-axis and around 4.5~cm$^{-1}$ for the $c$-axis. This feature is visible already at 30~K, much higher than the N\'eel temperature of 4.6~K, and it does not depend on external magnetic fields as high as 7~T. We conclude that this feature is independent of the antiferromagnetic order for UNi$_2$Al$_3$, and this might also be the case for UPd$_2$Al$_3$ and UPt$_3$, where a similar maximum in the optical conductivity was observed previously [M. Dressel \textit{et al}., Phys. Rev. Lett. {\bf 88}, 186404 (2002)]. 
\end{abstract}


\maketitle

\section{Introduction}
Heavy-fermion materials, such as UNi$_2$Al$_3$, are intermetallic compounds that contain $f$-electrons. They behave as metals, but when they are cooled below the coherence temperature, their properties change drastically due to the hybridisation between the delocalized conduction and localized $f$-electrons. Thus heavy fermions are model systems for strongly interacting electrons. The hybridisation leads to a gap in the density of states and a very large effective mass $m^*$. For UNi$_2$Al$_3$ the mass enhancement $m^*/m$ is estimated to be around 70. \cite{geibel1} Heavy-fermion behavior has been studied in detail for example by dc resistivity, susceptibility, and specific heat for many different compounds including the title compound UNi$_2$Al$_3$.\cite{geibel1,Pfleiderer} However, much less spectroscopic data are available.

\begin{figure}
	\centering
		\includegraphics{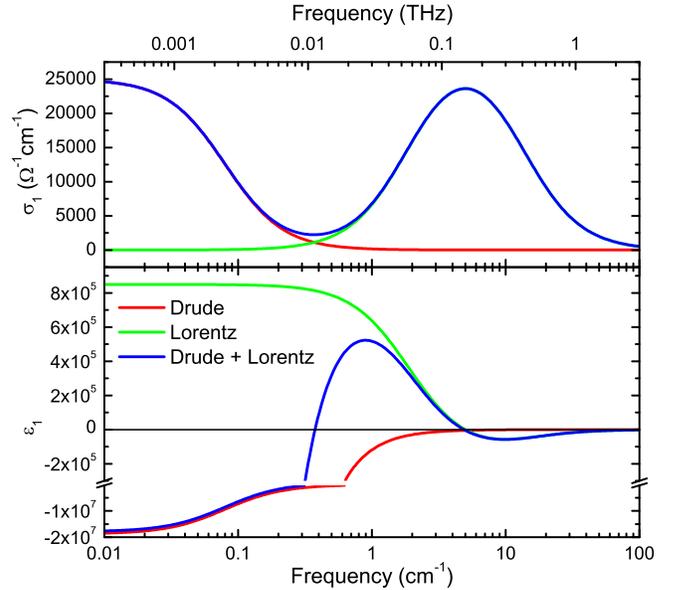}
	\caption{(Color online) The frequency dependence of the real part of the optical conductivity $\sigma_1$ and permittivity $\epsilon_1$ for Drude, Lorentz, and combined Drude and Lorentz response. The parameters used for the Drude behavior are $\sigma_{dc} = 25000$~$\Omega^{-1}{\rm cm}^{-1}$ and $\Gamma_D/(2\pi) = 0.08$~${\rm cm}^{-1}$. For the Lorentz oscillator we chose $\nu_0=\omega_0/( 2 \pi c) = 5$~${\rm cm}^{-1}$, $\nu_{P,L}=\omega_{P,L}/(2 \pi c) = 3800$~${\rm cm}^{-1}$, and $\Gamma_L = 11$~${\rm cm}^{-1}$.}
	\label{fig:skizze}
\end{figure}

Optical studies on heavy fermions are especially interesting because electromagnetic radiation couples directly to the electronic system, and the frequency of the radiation is an energy scale that can be adjusted to the material characteristics of interest.\cite{degiorgi,dressel} For temperatures above the coherence temperature, the real parts of the optical conductivity $\sigma_1(\omega)$ and the dielectric function $\epsilon_1(\omega)$ are expected to follow conventional Drude behavior:\cite{dressel} $\sigma_1(\omega) = \sigma_{dc} (1+\omega^2\tau_D^2)^{-1}$ and $\epsilon_1(\omega) = 1-\omega_{P,D}^2 (\omega^2+\frac{1}{\tau_D^2})^{-1}$. Here $\sigma_{dc}$ is the dc conductivity, $\omega_{P,D}= (4\pi\sigma_{dc}/\tau_D)^{1/2}$ the plasma frequency, $\tau_D$ the relaxation time, $\Gamma_D=1/(2\pi c \tau_D)$ the scattering rate, and $\omega$ the angular frequency. The frequency dependences of the real parts of $\sigma(\omega)$ and $\epsilon(\omega)$ in case of a Drude behavior are shown in Fig. \ref{fig:skizze}.

Below the coherence temperature, peculiar optical characteristics are known for heavy fermions. A renormalized Drude behavior at microwave frequencies was predicted by Millis and Lee\cite{millis} and found experimentally,\cite{awasthi, beyermann, tran, scheffler, SchefflerEPJB} also for UNi$_2$Al$_3$.\cite{scheffler2} Here the Drude conductivity has renormalized effective mass $m^*$ and scattering rate $\Gamma_D^*=1/(2 \pi c \tau_D^*)$. According to Millis and Lee,\cite{millis} these are related as $m^*/m = \Gamma_D / \Gamma_D^* = \tau_D^* / \tau_D$. This leads to an unaffected $\sigma_{dc} \propto \tau_D^* / m^*$ but renormalized scattering rate $\Gamma_D^*$ in the Drude behavior. Therefore heavy-fermion behavior is not apparent in the dc transport but clearly observable in the frequency dependence of the optical conductivity $\sigma(\omega)$.
The hybridisation gap in the infrared is another well-studied characteristic of heavy-fermion optics. \cite{dordevic,okamura} It is usually attributed to excitations over the gap in the density of states that develops due to the hybridisation,\cite{coleman} but there are also recent calculations that describe such a gap structure in the conductivity as a bandstructure effect.\cite{sichelschmidt,kimura} In the present study we concentrate on frequencies that are much lower than those typically assumed for the hybridization gap.

In the frequency range between 1~cm$^{-1}$ and 40~cm$^{-1}$, there are only few studies on heavy-fermion compounds. Some concentrate on the superconducting transition,\cite{Brown, Sudhakar} but only two heavy-fermion materials, namely UPd$_2$Al$_3$ \cite{dressel1,dressel2} and UPt$_3$,\cite{donovan} have been studied previously in a broad temperature range at these frequencies. For both compounds a maximum in the real part of the optical conductivity was discovered:\cite{dressel1,dressel2,donovan} for UPd$_2$Al$_3$ at 4~cm$^{-1}$ and for UPt$_3$ at 6~cm$^{-1}$. It can roughly be described by a Lorentzian oscillator with the following frequency dependence as sketched in Fig. \ref{fig:skizze}:
\begin{equation}
 \sigma_1(\omega) = \frac{\omega_{P,L}^2}{4\pi} \frac{\omega^2 / \tau_L}{(\omega_0^2 - \omega^2)^2 + \omega^2 / \tau_L^2}
\end{equation}
\begin{equation}
 \epsilon_1(\omega) = 1 +  \frac{\omega_{P,L}^2 (\omega_0^2 -\omega^2)}{(\omega_0^2 - \omega^2)^2 + \omega^2 / \tau_L^2}
\end{equation}
Here $\omega_0$ is the center frequency, $\omega_{P,L}$ the plasma frequency and $\Gamma_L=1/(2 \pi c \tau_L)$ the broadening of the Lorentzian oscillator.
In both materials the maximum in $\sigma_1$ develops at temperatures directly below the N\'eel temperature which is 14.5~K for UPd$_2$Al$_3$ and 5~K for UPt$_3$. This was the main reason to relate this feature to the antiferromagnetic ordering,\cite{dressel1,dressel2,grüner} but the details are not well understood. Thus it is important to study further heavy-fermion compounds in the frequency range between $\nu$ = 1~cm$^{-1}$ and 30~cm$^{-1}$ to examine whether this feature is generic for heavy-fermion compounds and to find out more about its origin.

UNi$_2$Al$_3$ is especially suited to address this question as it is very similar to the isostructural compound UPd$_2$Al$_3$.\cite{geibel1,geibel2} Both of them have a hexagonal crystal structure, and an anisotropy in the transport properties along the $a$- and $c$-axes.\cite{Jourdan,foerster} UPd$_2$Al$_3$ has a commensurable antiferromagnetic phase below 14.5~K with a magnetic moment of 0.85$\mu_B$.\cite{krimmel} For UNi$_2$Al$_3$ the phase transition to the antiferromagnetically ordered state takes place at $T_N$ = 4.6~K. Here the order is incommensurable with a magnetic moment of 0.24$\mu_B$.\cite{krimmel,Jourdan2} Furthermore, both compounds show a superconducting transition, UPd$_2$Al$_3$ at $T_c$ = 2~K, UNi$_2$Al$_3$ at 1~K.\cite{geibel1,geibel2} As the two compounds are very similar, we expect similar optical properties, and therefore there should be a maximum in the real part of the optical conductivity at a few wavenumbers for UNi$_2$Al$_3$ as well. An advantage of UNi$_2$Al$_3$ for these studies is that the antiferromagnetically ordered state can be suppressed with magnetic fields that can be combined with optical spectroscopy. Following the phase diagrams of UNi$_2$Al$_3$ determined for single crystals by S\"ullow \textsl{et al.}\cite{Süllow} and Tateiwa \textsl{et al.}\cite{tateiwa}, at 2~K the antiferromagnetically ordered state can be suppressed with a magnetic field of about 6~T aligned along the $a$-axis of UNi$_2$Al$_3$, whereas our own magnetoresistance studies on UNi$_2$Al$_3$ thin films (including the present film) revealed a somewhat higher critical field.\cite{SchefflerJPSJ}
For UPd$_2$Al$_3$ much higher fields of the order of 20~T are necessary to reach the phase boundary.\cite{grauel}

One further advantage of UNi$_2$Al$_3$ for our optical experiments is that the $a$- and $c$-axes lie within the plane of the thin film sample\cite{Jourdan,Jourdan1} and can be probed with incident light of different linear polarization; this allows us to conveniently study whether the anisotropy that is observed in the dc properties\cite{Jourdan} of UNi$_2$Al$_3$ is also present in the optical conductivity at frequencies of a few cm$^{-1}$.\cite{Ostertag} The optical properties of UNi$_2$Al$_3$ at higher frequencies, in the infrared, have been studied previously by Cao \textsl{et al.} for temperatures between 10~K and 300~K using a single crystal;\cite{cao} within the small spectral overlap above 30~cm$^{-1}$ their measurements are consistent with ours.

\section{Experimental Methods}\label{Experimental_Methods}
We studied a UNi$_2$Al$_3$ film epitaxially grown with molecular beam epitaxy by coevaporation of the constituent elements onto a heated, 1.044~mm thick YAlO$_3$(112) substrate. The (100)-axis of the thin film is perpendicular to the substrate surface.\cite{Jourdan1} For the transmission measurements at low frequencies, we need a very thin film, here 62~nm, to obtain a measurable transmission signal, and we need a large size of the sample, here 1~cm*1~cm, to avoid diffraction effects. The sample shows a superconducting transition at $T_c=0.46$~K and an antiferromagnetically ordered phase below $T_N=4.2$~K, as determined from the dc resistivity.\cite{SchefflerJPSJ} The residual resistivity ratio (RRR) is 5.5 along the $a$-axis. The substrate YAlO$_3$(112) is also anisotropic,\cite{scheffler1} and the main optical axes of the YAlO$_3$ substrate and the UNi$_2$Al$_3$ film, which are equal to the crystallographic axes, are tilted with respect to each other by 45$^\circ$.

\begin{figure}
	\centering
		\includegraphics{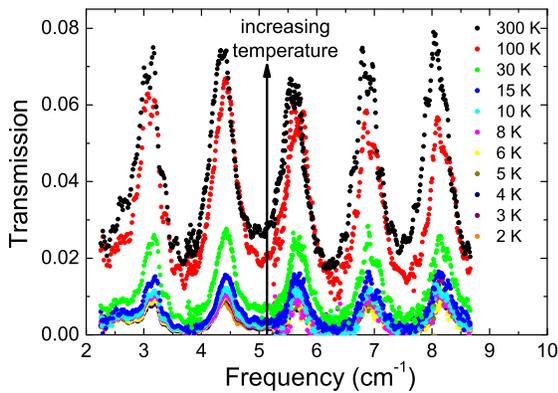}
	\caption{\label{fig:temperature_dependence}(Color online) Transmission spectra along the $a$-axis of the UNi$_2$Al$_3$ thin film on YAlO$_3$ substrate for different temperatures.}
\end{figure}

Transmission $Tr$ and phase shift $\phi$ of radiation passing through the sample were measured for frequencies from 2~cm$^{-1}$ to 40~cm$^{-1}$ with a Mach-Zehnder interferometer setup.\cite{gorshunov} The radiation sources were several backward wave oscillators with different frequency ranges. Temperatures between 2~K and 300~K were obtained using a home-built optical cryostat. Fig.~\ref{fig:temperature_dependence} shows the transmission spectra for the different temperatures and for frequencies below 9~cm$^{-1}$. The spectra show well pronounced Fabry-Perot resonances due to the dielectric YAlO$_3$ substrate. The overall transmission is strongly reduced with decreasing temperature, which directly indicates the increase of conductivity of the UNi$_2$Al$_3$ thin film.
The details of the analysis procedure, which has to go beyond the conventional approach\cite{dressel} because of the misaligned anisotropic substrate, are described in Appendix \ref{Analysis}.

\begin{figure}
	\centering
		\includegraphics{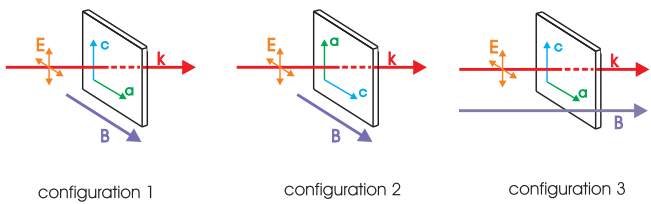}
	\caption{\label{fig:Konfigurationen} (Color online) Transmission measurements with an external magnetic field were performed in three different configurations of sample and magnetic field, each with two different polarizations of the electric field vector of the radiation.}
\end{figure}
			
Optical measurements in finite magnetic field up to 7~T were performed at temperatures down to 2~K using an optical cryostat by Oxford instruments which allows measurements in three different configurations with two polarizations of the radiation each, as shown in Fig.~\ref{fig:Konfigurationen}. In configuration 1, with the magnetic field aligned parallel to the $a$-axis of the UNi$_2$Al$_3$ thin film, a large influence of the magnetic field is expected \cite{Süllow}, and here we measured from $\nu =$~2~cm$^{-1}$ to 8.5~cm$^{-1}$ at $T =$~100~K, 10~K and 2~K. In configurations 2 and 3, we measured from $\nu =$~4~cm$^{-1}$ to 5.8~cm$^{-1}$ at $T =$~10~K and 2~K.

\section{Results}

\subsection{Optical conductivity}

\begin{figure*}
	\centering
		\includegraphics{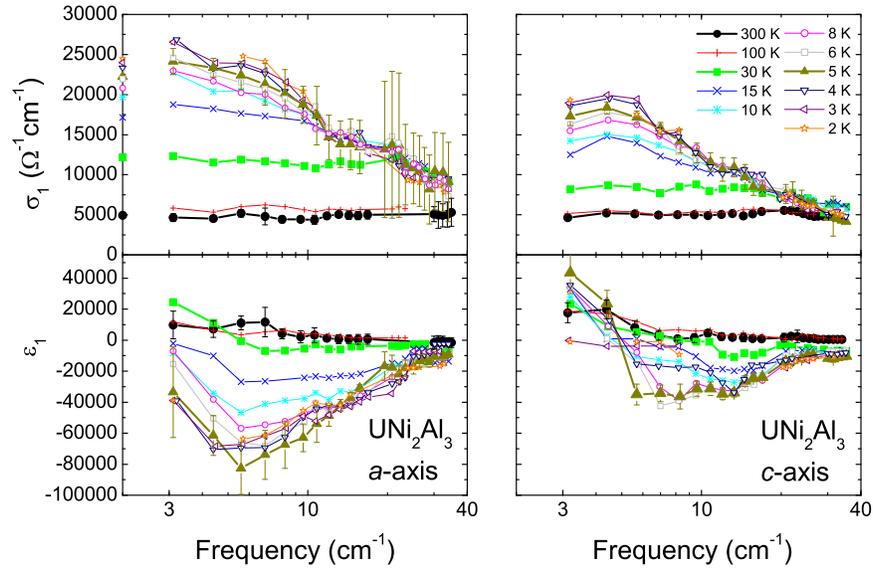}
	\caption{\label{fig:SpectraSigmaEps}(Color online) Real parts of optical conductivity and permittivity of UNi$_2$Al$_3$ along the $a$-axis (left) and $c$-axis (right) for different temperatures. Data points on the $\sigma_1$-axis indicate the dc conductivity.}
\end{figure*}

\begin{figure}[htbp]
	\centering
		\includegraphics{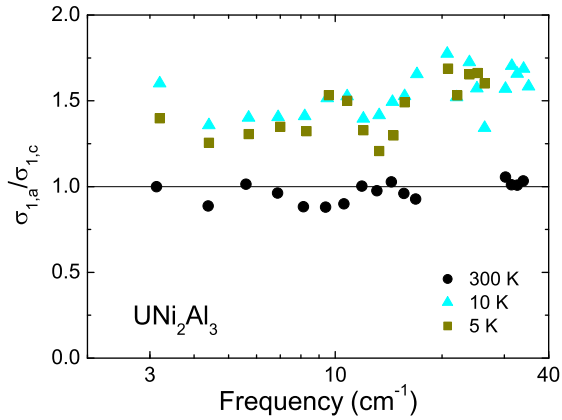}
	\caption{\label{fig:anisotropie}(Color online) Ratio of $\sigma_1$ along the $a$- and the $c$-axis for UNi$_2$Al$_3$, as a function of frequency for three exemplary temperatures.}
\end{figure}

\begin{figure}
	\centering
		\includegraphics{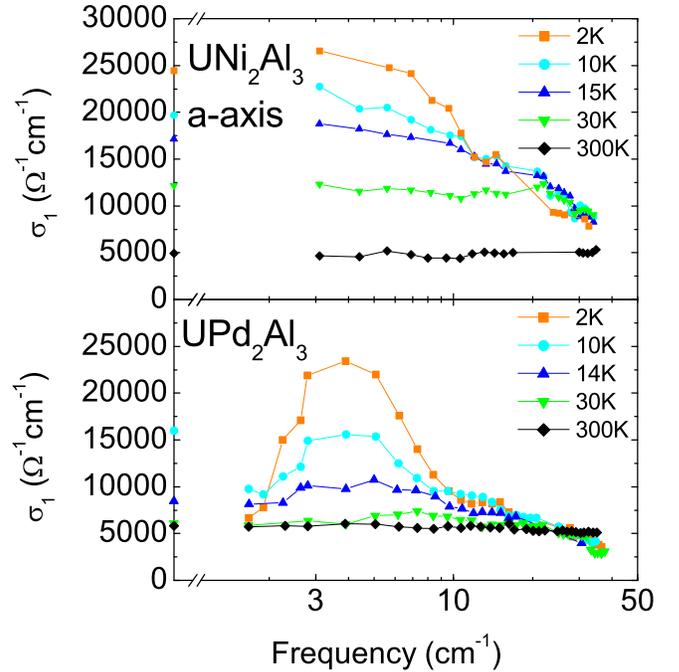}
	\caption{\label{fig:comparison_UNA_UPA}(Color online) The real part of the optical conductivity $\sigma_1$ for UNi$_2$Al$_3$ compared to the corresponding data for UPd$_2$Al$_3$ from literature.\cite{dressel2}}
\end{figure}

We determined the real parts $\sigma_1$, $\epsilon_1$ of the optical conductivity and permittivity of UNi$_2$Al$_3$ between 3~cm$^{-1}$ and 35~cm$^{-1}$ along $a$- and $c$-axes independently. In Fig.~\ref{fig:SpectraSigmaEps}, the resulting data are shown for both crystallographic axes. 
For $T$= 300~K and 100~K, $\sigma_1 \approx$ 5000~$\Omega^{-1}$cm$^{-1}$ is almost constant for both directions. This is expected since for temperatures above the coherence temperature the electrons should have a relaxation rate at infrared frequencies.
Below 30~K, $\sigma_1$ increases strongly with decreasing temperature. Furthermore, we observe the general trend that for our frequency range $\sigma_1$ decreases for increasing frequencies. But for the $c$-axis and the lowest accessible frequencies, a maximum in $\sigma_1$ can clearly be identified at a frequency of roughly 4.5~cm$^{-1}$.
With decreasing temperature the absolute value of the maximum rises, up to 20000~$\Omega^{-1}$cm$^{-1}$ at 2~K. Along the $a$-axis, $\sigma_1$ is 27000~$\Omega^{-1}$cm$^{-1}$ for 2~K at 3~cm$^{-1}$, the lower limit of our accessible frequency range. Here we do not observe yet a maximum in $\sigma_1$, but from microwave measurements \cite{scheffler} we know that $\sigma_1$ is smaller than 5000~$\Omega^{-1}$cm$^{-1}$ around 0.8~cm$^{-1}$ at these temperatures, i.e. there must be a maximum between 3~cm$^{-1}$ and 0.8~cm$^{-1}$. This is corroborated further if we examine the real part of the dielectric constant.
As shown schematically in Fig.~\ref{fig:skizze}, the maximum in $\sigma_1$ is connected in $\epsilon_1$ to a zero-crossing and a minimum at slightly higher frequencies. As evident from Fig.\ \ref{fig:SpectraSigmaEps}, we can clearly observe both, the minimum and the zero-crossing of $\epsilon_1$ for the $c$-axis. For the $a$-axis, we only observe the minimum in $\epsilon_1$ around 5~cm$^{-1}$, i.e. the zero-crossing of $\epsilon_1$ and the maximum in $\sigma_1$ have to be at slightly lower frequencies than accessed by our experiment. Therefore we conclude that the maximum in $\sigma_1$ is located close to 3~cm$^{-1}$ for the $a$-axis and 4.5~cm$^{-1}$ for the $c$-axis.

At temperatures below 30~K, an almost frequency-independent anisotropy appears. In Fig.~\ref{fig:anisotropie} the ratio of $\sigma_1$ along the $a$- and the $c$-axis shows a pronounced anisotropy of 50~\% for 5~K and 10~K. Thus the anisotropy at low temperatures that was already known from the dc transport properties\cite{Jourdan} is also present in the THz frequency range.\cite{Ostertag}

The error bars shown in Fig.~\ref{fig:SpectraSigmaEps} are determined by changing the parameters of the fit (to the Fabry-Perot resonances in the experimental data of $Tr$ and $\phi$) till there is a clear discrepancy between the fit and the raw data. Below 20~cm$^{-1}$, the error in $\sigma_1$ is smaller than 10~\%. Due to difficulties in the alignment above 20~cm$^{-1}$, the uncertainity in the phase shift gets bigger and this leads to a larger error in $\sigma_1$ and $\epsilon_1$ at higher frequencies. Additional confirmation for the data above 20~cm$^{-1}$ comes from independent reflectivity measurements which were performed on a much thicker UNi$_2$Al$_3$ film (150~nm thick) with a Fourier-transform spectrometer and which gave consistent results. 

When we compare the maximum in $\sigma_1$ of UNi$_2$Al$_3$ to the one known for UPd$_2$Al$_3$ from literature,\cite{dressel2} see Fig.~\ref{fig:comparison_UNA_UPA}, but also the one in UPt$_3$,\cite{donovan} it looks very similar. In all cases the maximum in the optical conductivity arises at low temperatures at frequencies between 3~cm$^{-1}$ and 7~cm$^{-1}$; it increases in strength with decreasing temperature and has an almost temperature-independent characteristic frequency. As all the compounds have similar properties, it is very likely that the origin for the maximum in the optical conductivity is the same in all three cases. However for UNi$_2$Al$_3$ there is an important difference compared to the others: for this compound, the maximum already sets in at 30~K. For the other two compounds the explanation for this feature was up to now connected to the antiferromagnetically ordered phase.\cite{donovan,dressel1,dressel2} As our UNi$_2$Al$_3$ sample orders antiferromagnetically only below 4.2~K, it cannot be explained in this picture why we clearly see the maximum already at 30~K. This suggests that the origin of the feature is not due to the antiferromagnetically ordered state for UNi$_2$Al$_3$, and then this might also not be the case for UPd$_2$Al$_3$ and UPd$_3$.
Moreover there is an interesting relation between the dc and optical measurements. If we compare the dc conductivity with the frequency dependent conductivity, as in Fig.~\ref{fig:SpectraSigmaEps}, it can be seen that within the accuracy of our measurement the value of the optical conductivity at 3~cm$^{-1}$ is equal to the dc conductivity along the $a$-axis. Furthermore, the anisotropy appears in dc as well as optical properties. Thus the THz conductivity seems to be closely linked to the dc conductivity.

\subsection{Optical properties in finite magnetic field}
\begin{figure}[htbp]
	\centering
		\includegraphics{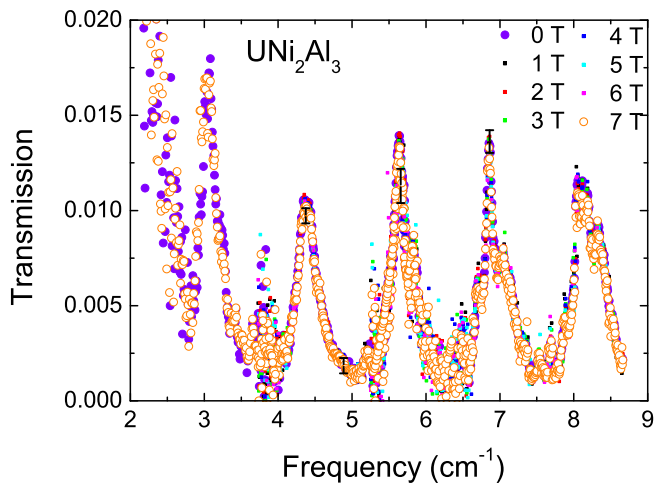}
	\caption{\label{fig:Feldabhängig}(Color online) Transmission spectra at 2~K with electric field polarized along the $a$-axis with external magnetic fields between 0~T and 7~T (see Fig.~\ref{fig:Konfigurationen}: configuration 1). The transmission shows no field dependence within the accuracy of our measurement.}
	\end{figure}

In order to clarify whether there is a connection between the observed optical feature and the antiferromagnetic state, we measured the transmission of the UNi$_2$Al$_3$ film in an applied magnetic field of up to 7~T in the different configurations shown in Fig.~\ref{fig:Konfigurationen}. According to the phase diagram determined on this particular sample,\cite{SchefflerJPSJ} we did not induce the transition to the paramagnetic phase with the highest field we applied in the optical experiments. But increasing the field up to 7~T at a temperature of 2~K, we considerably approached the phase boundary. Fig.~\ref{fig:Feldabhängig} shows the transmission spectra at 2~K for both the static magnetic field and the electric field of the THz radiation applied along the $a$-axis. Increasing the field from 0~T to 7~T, we could not detect any change of transmission within our (relative) accuracy of better than 10~\%. (This is in contrast to the temperature dependence, which we could easily detect, see Fig.~\ref{fig:temperature_dependence}.) Neither did we find any dependence on external magnetic field for the other configurations and temperatures that we tried.
If the THz feature were directly related to the antiferromagnetic phase, one would expect it to depend on magnetic field. The absence of any field dependence is a further indication that the observed feature is not connected to the antiferromagnetically ordered phase.
In the dc resistivity measurements on the same sample, we observed a magnetoresistace of only 4$\;\%$ for an applied field of 7~T.\cite{SchefflerJPSJ} From the accuracy of our transmission measurements, we cannot exclude that the THz conductivity in the measured frequency range changes by a similar, small amount. Thus, also the field-dependent behavior of the THz conductivity could match that of the dc conductivity.

\section{Conclusions}
We have determined the optical properties of a UNi$_2$Al$_3$ thin film at frequencies between 3~cm$^{-1}$ and 35~cm$^{-1}$ along $a$- and $c$-axes for the temperature range from 2~K to 300~K, and we have studied the magnetic-field dependence up to 7~T. In the optical conductivity we observe a maximum at low frequencies that appears below 30~K and grows with decreasing temperature. This seems to be the same feature that was already observed for UPd$_2$Al$_3$\cite{dressel1,dressel2} and UPt$_3$.\cite{donovan} From our study we can exclude that this feature is connected to the antiferromagnetically ordered phase as it sets in at temperatures much higher than the N\'eel temperature and cannot be influenced by fields as high as 7~T. This might also lead to a new interpretation for UPd$_2$Al$_3$ and UPt$_3$. Interestingly the absolute value of the maximum in the optical conductivity corresponds to the dc conductivity. 

For future studies it would be very interesting to perform field-dependent optical measurements with higher magnetic fields and higher resolution in the THz range. If one can reach the field-induced phase transition without observing any change in the optical properties, this would definitely rule out a connection of the observed feature to the antiferromagnetic state. 
To determine whether this feature is characteristic for all heavy-fermion compounds or not, the optical conductivity of a non-uranium-based heavy fermion and a heavy fermion without antiferromagnetic order should be studied. 
Considering the recent improvements in the growth of Ce-based heavy-fermion thin films, this might become feasible in the future.\cite{Shishido}

\begin{acknowledgments}
We thank Boris Gorshunov for his support during THz measurements and discussions, and Stefan Kaiser for additional infrared measurements. Financial support by the German Research Foundation (DFG) is acknowledged. 
\end{acknowledgments}

\appendix
\section{Analysis for anisotropic samples}\label{Analysis}
\begin{figure}
	\centering
		\includegraphics{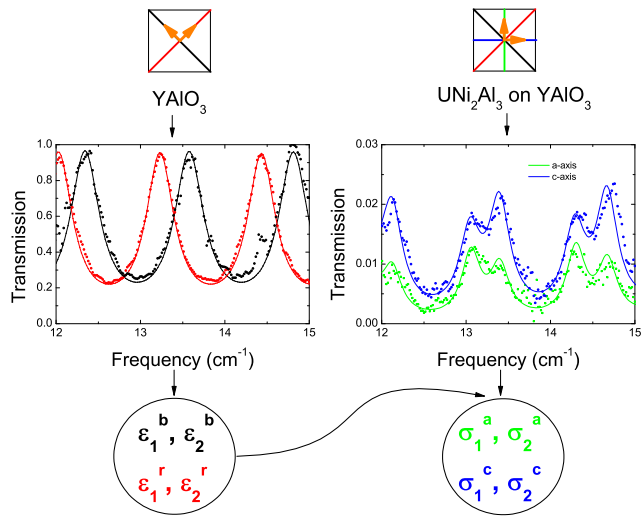}
	\caption{\label{fig:Auswertung-skizze}(Color online) On top, schematics of the empty substrate YAlO$_3$ with the main axes in red and black and the UNi$_2$Al$_3$ film on YAlO$_3$ with the main axes of the film in green and blue and the main axes of the underlying substrate in red and black are shown. The orange arrows indicate the polarization of the light during the measurements. The transmission of the empty YAlO$_3$ substrate along its main axes shows usual Fabry-Perot resonances. By fitting transmission and phase shift (here only transmission is shown) we can determine the optical parameters along the ''black'' main axis ($\epsilon_1^b,\epsilon_2^b$) and along the ''red'' main axis ($\epsilon_1^r,\epsilon_2^r$). These parameters are used in the fit of transmission and phase shift of the UNi$_2$Al$_3$ film on YAlO$_3$ to extract $\sigma_1$ and $\sigma_2$ (or equivalently $\epsilon_1$) of UNi$_2$Al$_3$ along $a$- and $c$-axis. The double maxima in the transmission of UNi$_2$Al$_3$ film on YAlO$_3$ are caused by the birefringent substrate because the radiation is not polarized along any of its main axes.}
\end{figure}
 
\begin{figure}
	\centering
		\includegraphics{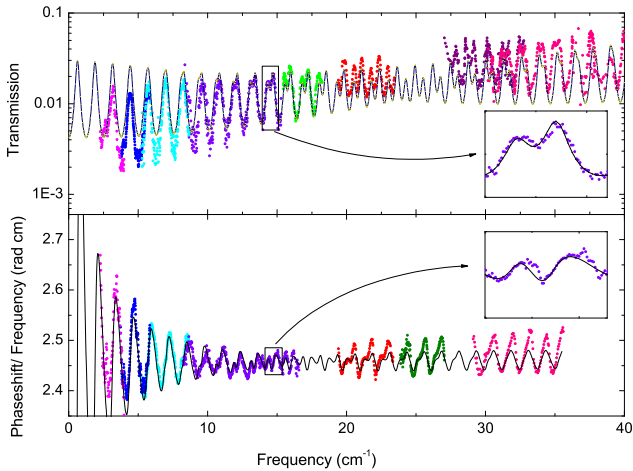}
	\caption{\label{fig:TrPh_UNAblue_4K}(Color online) Transmission and phaseshift of a UNi$_2$Al$_3$ film on a YAlO$_3$ substrate at 4~K along the $c$-axis of UNi$_2$Al$_3$. The dotted spectra are measured, and the solid line is a fit optimized around 14~cm$^{-1}$. This frequency range is enlarged in the insets. By fitting each Fabry-Perot maximum separately we find the frequency dependence of the optical properties.}
\end{figure}

Based on the Fresnel formula, the analysis of Transmission $Tr$ and phaseshift $\phi$ data for isotropic samples is well established:\cite{dressel,gorshunov} due to the finite thickness of the dielectric substrate, there are multiple reflections of the radiation in the substrate, and we observe Fabry-Perot resonances. In the analysis procedure we simultaneously fit transmission and phase shift for each Fabry-Perot transmission maximum to determine the frequency dependence of optical parameters. First we analyze the data of an empty reference substrate and determine the optical parameters of the substrate material. When we analyze the data of the UNi$_2$Al$_3$ sample, we use these optical parameters of the substrate to determine the parameters of the film material, see Fig.~\ref{fig:Auswertung-skizze}.

Compared to the conventional case, with our sample we have the rather special case of an anisotropic film on an anisotropic substrate with the main axes tilted by 45$^\circ$ with respect to each other. Anisotropic samples can only be described in the traditional way when the radiation is polarized along one of the main optical axes of the substrate. To resolve the anisotropy of the conductivity of the UNi$_2$Al$_3$, the measurements on the UNi$_2$Al$_3$ sample have to be performed with polarization along the main axes of the UNi$_2$Al$_3$ film (which are misaligned with respect to the substrate main axes). But we perform the reference measurements on the empty substrate with polarization along the main axes of the YAlO$_3$, see Fig.~\ref{fig:Auswertung-skizze}. To be able to describe the bare anisotropic substrate for any direction of polarization, we developed an extended analysis procedure.\cite{scheffler1} The main idea is to split the electric field vector of the incoming light into its projections along the main axes. Then we have to determine how the total measured transmission or phase shift is composed of the transmission and phase shift of radiation polarized along the two substrate main axes. When the incident radiation is not polarized along the main axes, the transmission shows characteristic double maxima (see Figs.~\ref{fig:Auswertung-skizze} and \ref{fig:TrPh_UNAblue_4K}). Transmission and phase shift for substrate plus film were measured with polarization aligned to the main axes of the UNi$_2$Al$_3$ film; thus we directly and individually obtain the response of the film along each of the two crystallographic axes, $a$ and $c$. For these measurements it is crucial that the radiation passes first the film and then the substrate as the polarization is changed in the anisotropic substrate.

In Fig.~\ref{fig:TrPh_UNAblue_4K} a typical transmission and phase spectrum is shown together with a fit optimized for 14~cm$^{-1}$. The characteristic double maxima structure is due to the anisotropic substrate. In the insets of Fig.~\ref{fig:TrPh_UNAblue_4K}, the frequency range around 14~cm$^{-1}$ is shown in detail with the Fabry-Perot maximum for which this particular fit was determined. Here we want to point out that the peculiar feature which is the main topic of this article (maximum in $\sigma_1$), is already evident from the raw data: in Fig.~\ref{fig:TrPh_UNAblue_4K}, it shows up as the Fabry-Perot maximum around 4.5~cm$^{-1}$ with the absolute value substantially suppressed compared to the adjacent maxima.


\end{document}